\documentclass[a4paper,twocolumn,preprintnumbers,citeautoscript,newabstract,aps]{revtex4}
\usepackage[utf8]{inputenc}
\usepackage{amsmath,amsfonts,amssymb,graphics}
\usepackage[normalem]{ulem}
\usepackage{graphicx}
\usepackage{units}
\usepackage{mathrsfs}
\usepackage{bbm}
\usepackage{pifont}
\usepackage{braket}
\usepackage{units}
\usepackage{footmisc}
\usepackage[dvipsnames]{xcolor}
\usepackage{mathtools}
\usepackage{xspace}
\usepackage{booktabs}

\usepackage[bookmarks=false,hyperfootnotes=false]{hyperref}

\definecolor{nicered}{rgb}{0.5,.0,.0}
\definecolor{darkblue}{rgb}{0,.1,.9}
\definecolor{lightblue}{rgb}{0,.1,.6}
 \definecolor{darkgreen}{rgb}{0.0,0.2,0.0}
\hypersetup{colorlinks, citecolor=darkgreen ,linkcolor=darkblue, urlcolor=lightblue}

\newcommand{\Eqref}[1]{Eq.~\eqref{#1}}
\newcommand{\Figref}[1]{Fig.~\ref{#1}}

\newcommand{\keV}{\ensuremath{\,\mathrm{keV}}\xspace}

\newcommand{\GeV}{\ensuremath{\,\mathrm{GeV}}\xspace}
\newcommand{\eV}{\ensuremath{\,\mathrm{eV}}\xspace}
\newcommand*{\rep}[2][]{\ensuremath{{\boldsymbol{#2}#1}}}

\newcommand{\I}{\mathrm{i}}

\newcommand{\SU}[1]{\ensuremath{\mathrm{SU}(#1)}}
\newcommand{\U}[1]{\ensuremath{\mathrm{U}(#1)}}

\newcommand{\Up}{\ensuremath{\mathrm{U}(1)_{\mathrm{X}}}\xspace}
\newcommand{\N}[1]{\ensuremath{N_{#1}}}
\newcommand{\Nb}[1]{\ensuremath{\overline{N}_{#1}}}
\newcommand{\vH}{\ensuremath{v_h}\xspace}
\newcommand{\vP}{\ensuremath{v_{\phi}}\xspace}
\newcommand{\vS}{\ensuremath{v_s}\xspace}
\newcommand{\vbar}{\ensuremath{\bar{v}}\xspace}
\newcommand{\Zp}{\ensuremath{Z'}\xspace}
\newcommand{\hS}{h_{S}\xspace}

\newcommand{\eM}{\ensuremath{\varepsilon}\xspace}
\newcommand{\mZp}{\ensuremath{m_{\Zp}}\xspace}
\newcommand{\Mpl}{\ensuremath{M_{\mathrm{Pl}}}}

\newcommand{\myvec}[1]{\ensuremath{\begin{pmatrix}#1\end{pmatrix}}}

\definecolor{darkgreen}{rgb}{0.0, 0.6, 0.2}

\allowdisplaybreaks[1]

\begin{document}


\title{\textbf{\boldmath \Large Neutrino self-interactions and XENON1T electron recoil excess \unboldmath}}

\author{Andreas Bally}
\email[]{andreas.bally@mpi-hd.mpg.de}
\author{Sudip Jana}
\email[]{sudip.jana@mpi-hd.mpg.de}
\author{Andreas Trautner}
\email[]{trautner@mpi-hd.mpg.de}
\affiliation{\vspace{0.2cm}Max-Planck-Institut f\"ur Kernphysik, Saupfercheckweg 1, 69117 Heidelberg, Germany}
%

\begin{abstract}
The XENON1T collaboration recently reported an excess in electron recoil events 
in the energy range between $1-7\keV$.
This excess could be understood to originate from the known solar neutrino flux,
if neutrinos couple to a light vector-mediator with strength $g_{\nu N}$ that kinetically mixes with the photon with
strength $\chi$, and $g_{\nu N}\chi\sim10^{-13}$.
Here, we show that such coupling values can naturally arise in a renormalizable model of long-range vector-mediated 
neutrino self-interactions. 
The model could be discriminated from other explanations of the XENON1T excess by the characteristic $1/T^2$ energy dependence of 
the neutrino-electron scattering cross section. 
Other signatures include invisible Higgs and $Z$ decays and lepto-philic charged Higgses at a few 100\,GeV. ALPS~II will 
probe part of the viable parameter space.
\end{abstract}

\maketitle\widowpenalty10000\clubpenalty10000
A recently publicized search for low-energy electronic recoil events performed with the XENON1T detector 
yielded some unexpected excess over background, statistically significant at the level of $3.2\sigma$~\cite{Aprile:2020tmw}.
New physics explanations of this excess could include axions produced in the Sun~\cite{Aprile:2020tmw}, 
an unexpectedly sizable neutrino magnetic moment~\cite{Aprile:2020tmw}, inelastic semi-annihilation recoils \cite{Smirnov:2020zwf}, 
axion-like particle warm dark matter~\cite{Takahashi:2020bpq}, or a fast Dark Matter component~\cite{Kannike:2020agf}.
The solar axion and neutrino magnetic moment explanations 
are already practically ruled out, respectively by stellar cooling~\cite{Giannotti:2017hny},
or white dwarfs~\cite{Corsico:2014} and globular cluster cooling constraints~\cite{Arceo-Diaz:2015pva, Diaz:2019kim}.
A more conventional origin of the excess would be an unaccounted Tritium background 
in the detector~\cite{Aprile:2020tmw}, or simply a statistical fluctuation.
Notwithstanding this perception, 
here we proclaim a new physics explanation of the excess 
based on a UV-complete neutrino self-interaction model recently proposed by M.\,Berbig and
two of the authors~\cite{Berbig:2020wve}.

Elastic scattering of solar neutrinos off electrons is a subdominant background 
for low-energy electron recoil events in XENON1T with around $220$ expected events~\cite{Aprile:2020tmw}.
The inclusion of a light mediator that couples neutrinos to electrons could change this
conclusion and explain the observed excess, just as in the case of an enhanced neutrino magnetic moment.
However, generic light mediators, especially when coupled to electrons,
face severe constraints, see e.g.~\cite{Redondo:2015iea, Gherghetta:2019coi, Fabbrichesi:2020wbt}.
As in \cite{Berbig:2020wve}, we circumvent the most rigorous constraints
from cosmology by relying on a low temperature phase transition in the 
neutrino sector~\cite{Vecchi:2016lty}. 
However, different from \cite{Berbig:2020wve}, where mediator masses of $\mZp\sim\mathcal{O}(10)\eV$
were considered, we here focus on a parameter region with $\mZp\lesssim10^{-4}\eV$.
The reason is that in order to accommodate the observed excess at XENON1T there needs to be a
sufficiently strong interaction between electrons and the new mediator, 
which is excluded for the former region. 
In the new parameter region, the phase transition after which neutrinos start mixing with the hidden sector 
happens after recombination, i.e.\ below $T\sim1\eV$. 

\begin{figure}[t]
\centering\includegraphics[width=1\linewidth]{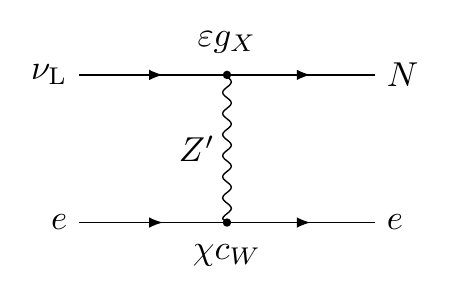}
\caption{\label{fig:nuNee}
Feynman diagram for inelastic neutrino-to-hidden-neutrino up-scattering on electrons 
mediated by a light \Zp. 
The couplings are generated by active-hidden neutrino mixing on the one side, and $\Up-\U1_\mathrm{Y}$ kinetic-mixing on the other.}
\end{figure}
The UV-complete and renormalizable model has been presented in some detail in \cite{Berbig:2020wve}
and we will only highlight the most important features here.
For the present analysis, relevant terms of the low energy effective Lagrangian in the mass basis read
\begin{equation}\label{eq:Leff}
 \mathcal{L}_\mathrm{eff}~=~\eM\,g_X\,\Zp_\mu\,\Nb{}\gamma^\mu\nu_\mathrm{L} + \chi\,c_W\,\Zp_\mu\,\mathcal{J}^\mu_{\mathrm{e.m.}}\;.
\end{equation}
Here, \Zp is the gauge boson of a new \Up gauge symmetry with coupling $g_X$, 
$N$ and $\nu_\mathrm{L}$ denote respectively the new hidden and left-handed SM neutrinos while \eM is the neutrino-hidden-neutrino mixing, 
$c_W$ is the cosine of the electroweak mixing angle~\footnote{%
We abbreviate trigonometric functions of all angles by $\sin\theta_i\equiv s_i$, $\cos\theta_i\equiv c_i$, and $\tan\theta_i\equiv t_i$ 
in this work.},
$\chi$ the strength of gauge-kinetic mixing of \Up and $\U1_{\mathrm{Y}}$ induced by an operator $\mathcal{L}_\chi=-(s_\chi/2)B^{\mu\nu}X_{\mu\nu}$~\cite{Galison:1983pa,Holdom:1985ag}, 
and $\mathcal{J}^\mu_{\mathrm{e.m.}}$ the standard electromagnetic current.
We only state \eqref{eq:Leff} for a single flavor of neutrinos $\nu_\mathrm{L}$ here, which is to be understood as a template for solar electron-neutrinos, 
while the extension to other flavors is straightforward.

\begin{figure}[t]
\centering\includegraphics[width=1\linewidth]{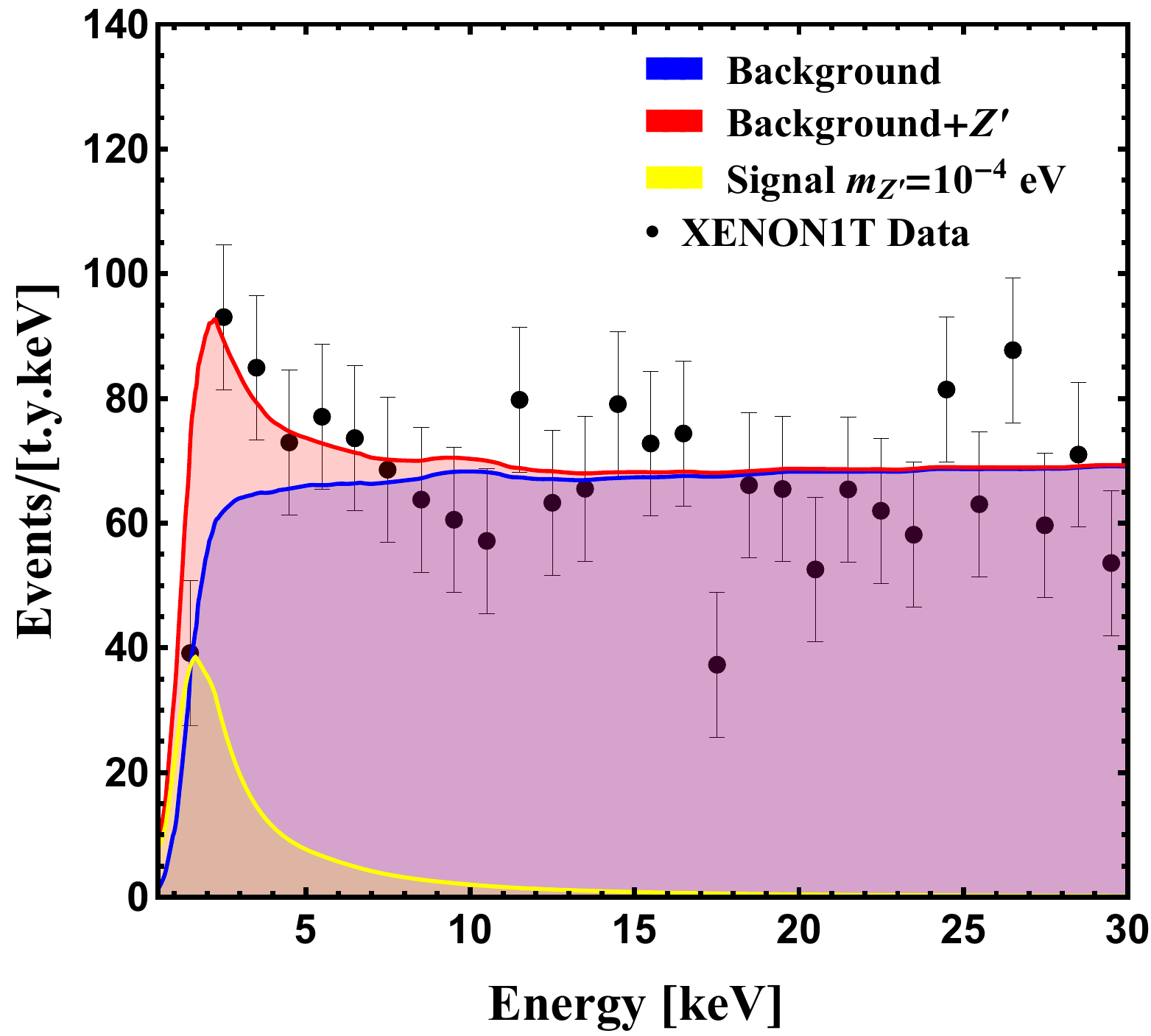}
\caption{\label{fig:spectrum}
Event rate for electronic recoils as a function of the electron recoil energy at XENON1T in our model. 
Data and background taken from \cite{Aprile:2020tmw}.}
\end{figure}
From \eqref{eq:Leff} it is possible to inelastically up-scatter neutrinos to hidden neutrinos on 
electrons in the XENON1T detector, see Fig.~\ref{fig:nuNee} (there is also elastic neutrino electron scattering, but this is suppressed
by another insertion of~\eM).
Taking $\mZp\ll \sqrt{2m_e T}$ and neglecting neutrino masses, the differential cross section for this process is given by (see e.g.\ \cite{Lindner:2018kjo})
\begin{equation}\label{eq:UpScatCS}
 \frac{\mathrm{d}\,\sigma_{\nu e\rightarrow N e}}{\mathrm{d}T}~=~ \frac{\eM^2\,g_X^2\,\chi^2\,c_W^2}{16\,\pi\,m_e\,T^2}\left[ 1+\left(1-\frac{T}{E_\nu}\right)^2-\frac{m_e\,T}{E_\nu^2} \right].
\end{equation}
Here $T$ is the electron recoil energy, $E_\nu$ the incident neutrino energy and $m_e$ the electron mass.
Note the $1/T^2$ enhancement of the differential cross section 
at low recoil energies. This is akin to the $1/T$ enhancement in the scattering induced by a neutrino magnetic moment.
With more data it should become possible to discriminate between the energy dependence of our model and a 
magnetic moment scattering, for example. 
Also note that without assuming $\mZp\ll \sqrt{2m_e T}$, the actual form of the propagator relevant 
in \eqref{eq:UpScatCS} is $(2m_e T+m_{Z^{\prime}}^2)^2$ and there is no enhancement for low-energy recoils,
which practically rules out an explanation of the excess for mediators heavier than $\sqrt{m_e T}$.

The low-energy solar neutrino flux consists essentially of the continuous pp and discrete ${}^{7}\mathrm{Be}$ flux components~\cite{Bahcall:2004mz},
\begin{align}
 \phi_{\mathrm{pp}}~=&~5.94\times 10^{10} \mathrm{cm^{-2}s^{-1}}\;, \\
 \phi_{{}^{7}\mathrm{Be}}~=&~4.86\times 10^{9} \mathrm{cm^{-2}s^{-1}}\;.
\end{align}
These neutrinos are affected by vacuum-dominated flavor oscillations  
resulting in a survival probability of \cite{Robertson:2012ib}
\begin{equation}
    P_{ee}~=~\cos^4\theta_{13}\left(1-\frac{1}{2}\sin^2\,2\theta_{12}\right)+\sin^4\theta_{13}\;,
\end{equation}
for electron neutrinos arriving at Earth.

To compute our signal prediction, we take into account XENON1T detection and selection efficiency \cite{Aprile:2020tmw} $\epsilon(T)$
as well as the finite detector energy resolution by a gaussian smearing \cite{Aprile:2017aty,Aprile:2020yad}.
Similarly to the XENON1T analysis we use the Free Energy Approximation
\begin{equation}
   \frac{\mathrm{d}\,\sigma_{\mathrm{tot}}}{\mathrm{d}T}=\sum_{i=1}^{54}\Theta(T-B_i)\frac{\mathrm{d}\,\sigma_{\nu e\rightarrow N e}}{\mathrm{d}T}\;,
\end{equation}
to take into account Xenon electron binding energies. At $\keV$ energies this is a good approximation 
to more sophisticated computations \cite{Chen:2016eab,Hsieh:2019hug}.
The differential event rate is then computed by the convolution
\begin{widetext}
\begin{equation}
    \frac{\mathrm{d}\,N(T_r)}{\mathrm{d}T_r}=
    N_0\times t \times 
    \int \mathrm{d}T \mathrm{d}E_{\nu}\, \frac{\mathrm{d}\,\phi (E_\nu)}{\mathrm{d} E_\nu}\,P_{ee}\,
    \frac{\mathrm{d}\,\sigma_{\mathrm{tot}}}{\mathrm{d}T}\,\Theta\left(\frac{2E_\nu^2}{m_e+2E_\nu}-T\right)\,\epsilon(T_r)\,g^{\mathrm{Gauss}}(T_r,T)\;,
\end{equation}
\end{widetext}
where $T$ and $T_r$ are the actual and reconstructed electron recoil energies, respectively.

Fitting this to the observed excess we obtain Fig.~\ref{fig:spectrum}.
The best fit point has 
\begin{equation}\label{eq:couplings}
 \eM\,g_X\,\chi~=~\left(2.0^{+0.7}_{-0.9}\right)\times10^{-13}\;\quad \left(95\%\mathrm{C.L.}\right)\;,
\end{equation}
and is statistically preferred over the background-only hypothesis by $3\sigma$.

Note that nothing in our analysis prevents us from considering flavors other than electron-neutrinos
in \eqref{eq:Leff}. This allows the possibility that also the subdominant non-electron-flavor solar neutrino flux
contributes to the excess. 
If the absolute relevant flux changes by a factor $f$, it is straightforward see that our result in \eqref{eq:couplings}
should be rescaled by $1/\sqrt{f}$.

\medskip
We now introduce our complete model in which $\mathcal{L}_\mathrm{eff}$ and the parameter region \eqref{eq:couplings} 
is naturally obtained.
New particles and their charges under the new \Up gauge symmetry are shown in Tab.~\ref{tab:model}.
We introduce a pair of SM-neutral but \Up charged chiral fermions $\N{1,2}$
and two new scalars $\Phi$ and $S$~\footnote{%
Similar models albeit in a completely different range of parameters have been conceived 
in \cite{Farzan:2016wym,Farzan:2017xzy,Denton:2018dqq} (see also \cite{Ballett:2019pyw,Ballett:2019cqp} 
for similar but somewhat incomplete models).}.
New interaction terms for the SM lepton doublet $L=(\nu_\mathrm{L},e_\mathrm{L})^\mathrm{T}$ are given by
\begin{equation}\label{eq:Lint}
\mathcal{L}_\mathrm{\mathrm{new}} = -y\,\bar{L}\,\tilde{\Phi}\,\N1-M\,\N1\,\N2 + \mathrm{h.c.}\;,
\end{equation}
where $\tilde{\Phi}:=\mathrm{i}\sigma_2\Phi^*$, $y$ is a dimensionless Yukawa coupling,
and $M$ has mass-dimension one.
We only discuss the one-generation case here, with the extension to three generations of SM leptons and 
multiple generations of hidden fermions being straightforward.
\renewcommand{\arraystretch}{1.5}
\begin{table}[t]
\begin{center}
\begin{tabular}{lccccc}
  \toprule[1pt]
  Field & $\Phi$ & $\N1$ & $\N2$ & $S$ & $X_\mu$ \\
  \hline 
  $\SU2_{\mathrm{L}}\times\U1_{\mathrm{Y}}$ & $(\rep{2},\frac12)$ & $\emptyset$ & $\emptyset$ & $\emptyset$ & $\emptyset$ \\
  \Up & $+1$ & $+1$ & $-1$ & $+1$ & $0$ \\
  \bottomrule[1pt]
  \end{tabular}
  \end{center}
  \caption{\label{tab:model} 
  New fields and their charges under SM and new $\Up$ gauge symmetry.}
\end{table}
We consider the most general possible scalar potential, cf.~\cite{Berbig:2020wve} for details,
and decompose the scalars as 
\begin{align}
 &H=\myvec{h^+ \\ \frac{1}{\sqrt{2}}\left(h + \I a_h  \right)},\quad 
 \Phi=\myvec{\phi^+ \\ \frac{1}{\sqrt{2}}\left(\phi + \I a_{\phi} \right)}, \\
 &\omit\hfill\text{and} $\quad S=\frac{1}{\sqrt{2}}\left(s + \I a_s\right)\;.$ \hfill
\end{align}
We assume that all neutral scalars obtain vacuum expectation values (VEVs) \mbox{$v_\sigma:=\langle \sigma\rangle$} for \mbox{$\sigma=h,\phi,s$}.
$\vH$ spontaneously breaks EW symmetry, $\vS$ breaks \Up, 
while $\vP$ breaks both. 
We will always assume the hierarchy $\vH\ggg\vS,\vP$, which is required by 
the assumption of a light $\Zp$.
Next to the SM Higgs boson $H$, 
the physical scalar spectrum consists of a pair of heavy ($>100\GeV$) charged scalars $\Phi^\pm$,
a pair of mass-degenerate heavy neutral scalar and pseudo-scalar $\Phi$ and $A$ ($|m_{\Phi^\pm}-m_{\Phi}|\lesssim 120\,\GeV$ by electroweak precision),
as well as a sub-keV light scalar $\hS$, 
all with very lepton-specific couplings and practically no coupling to quarks.

The photon is exactly the same massless combination of EW bosons as in the SM. 
The very SM-like $Z$ boson contains a miniscule admixture of the new gauge boson~$X$,
\begin{align}
Z_\mu &= c_X\left(c_W\,W^3_\mu - s_W\,B_\mu\right) + s_X\,X_\mu\;,
\end{align}
with an angle 
\begin{equation}\label{eq:sinX}
 s_X\approx-2\,c_W\frac{g_X}{g_2}\left(\frac{\vP}{\vH}\right)^2\lll1\quad\text{and}\quad c_X\approx1\;.
\end{equation}
Gauge-kinetic mixing from the operator $\mathcal{L}_\chi$ shifts the \Zp coupling to the 
SM neutral current only by a negligible amount proportional to 
$\chi\,\mathcal{O}(\mZp^2/m_Z^2)$~\cite{Baumgart:2009tn, Babu:2017olk} (given $\mZp\ll m_Z,\chi\ll1$). 
However, it will introduce the coupling of \Zp to the electromagnetic current shown in Eq.~\eqref{eq:Leff},
which is instrumental for our explanation of the XENON1T excess.
The masses of the physical neutral gauge bosons are
\begin{equation}
 m_Z\approx\frac{g_2\,\vH}{2c_W}\quad\text{and}\quad m_{\Zp}\approx g_X\sqrt{\vP^2+\vS^2} =: g_X\,\vbar\;.
\end{equation}
After $\phi$ assumes its VEV $\vP$, the Yukawa coupling $y$ will introduce 
bi-linear mixing between SM neutrinos and the new hidden neutrinos. 
For exactly massless neutrinos this implies a mixing of $\nu_\mathrm{L}$
and $\Nb2$ by an exact angle
\begin{equation}\label{eq:eM}
\tan \eM = (y\vP)/(\sqrt{2}M)\;.
\end{equation}
\begin{figure}[t]
\centering\includegraphics[width=1\linewidth]{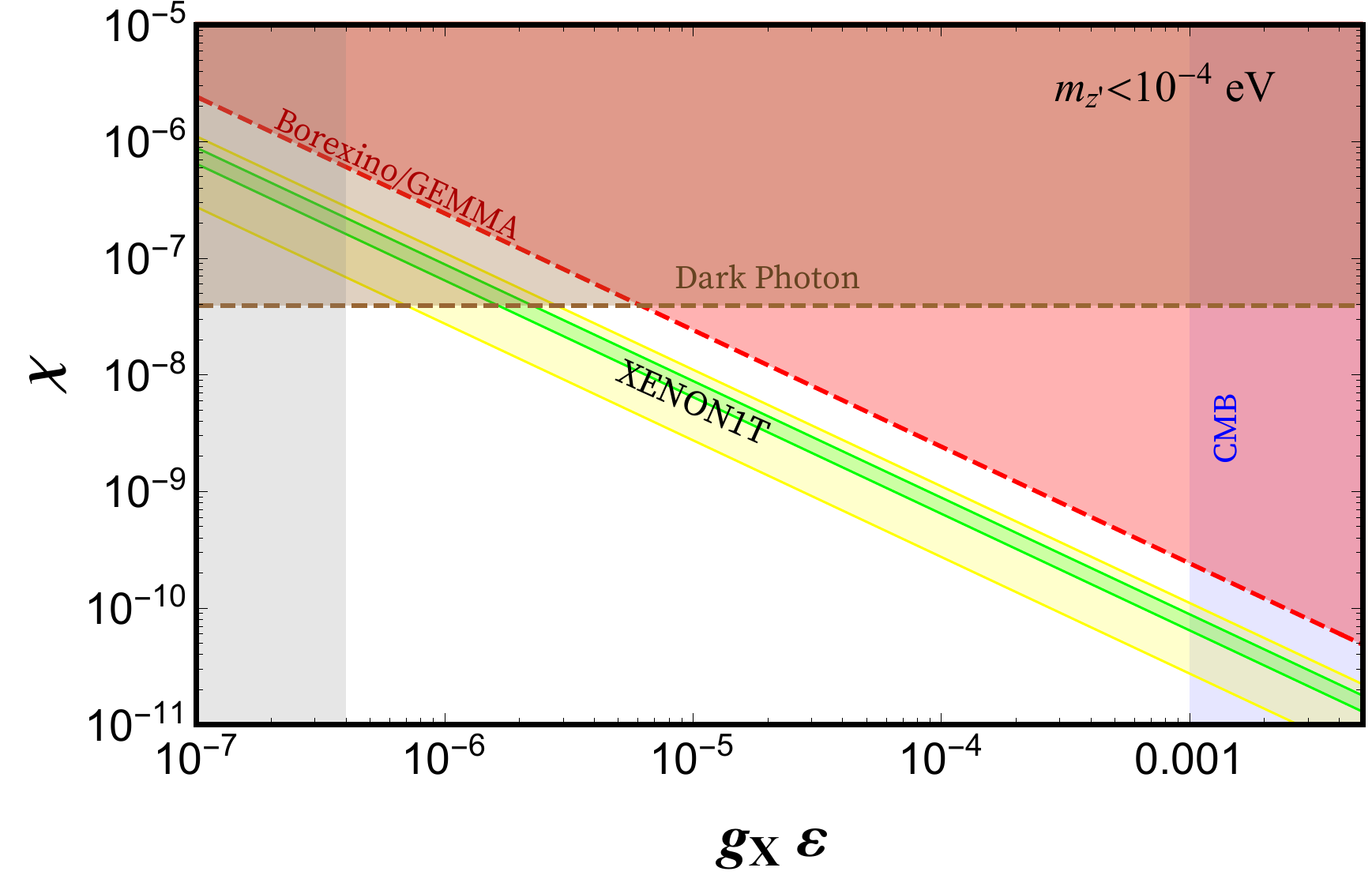}
\caption{\label{fig:money}
Constraints on electron neutrino scattering in the $\eM g_X$-$\chi$-plane.
Borexino/GEMMA limits \cite{Agostini:2018uly,Beda:2013} and XENON1T signal depends 
directly on $\eM g_X \chi$. 
Direct limits on $\chi$ arise from Dark Photon searches \cite{Redondo:2015iea, Gherghetta:2019coi, Fabbrichesi:2020wbt}, 
and they are essentially independent of $\mZp$ as long as $\mZp\lesssim 10^{-4}\mathrm{eV}$ (they
only get weaker below $10^{-9}\eV$). 
Constraints on $\eM^2 g_X$ are given by Eq.~\eqref{eq:gNu}, with the upper end (blue) excluded by neutrino free-streaming during CMB.
Their appearance in the plot slides with the exact choice of $\eM$; and we display the widest limit possible for $\eM$
constrained as in Eq.~\eqref{eq:epsilon}.}
\end{figure}
Depending on the specific neutrino mass generation mechanism (see for example \cite{Mohapatra:1986su,Bertuzzo:2018ftf, Dev:2012sg} for
mechanisms compatible with the model), and in particular whether or not there is violation of lepton number,
this will slightly change for massive neutrinos in which case there can also be a slight admixture of $\Nb1$
and \eqref{eq:eM} becomes approximate. In any case, $N$ in Eq.~\eqref{eq:Leff} should be understood as the resulting hidden-neutrino mass 
eigenstate which will have a (Dirac-)mass $M_N\approx(M^2+ y^2 \vP^2/2)^{1/2}$.
By this mixing, SM neutrinos pick up a coupling to \Zp from the gauge coupling of $\N{1,2}$.
This gives rise to the first term in \eqref{eq:Leff} but also to a pure SM neutrino \Zp interaction
proportional to $g_X\eM^2$. For temperatures $T\gg m_{\Zp}$, while neutrino mixing
via $\eM$ is relevant, \Zp will be effectively massless giving rise to an 
induced long-range four-neutrino interaction with thermally averaged rate
$\Gamma\sim\eM^8 g_X^4\,T$. Requiring this rate not to surpass the Hubble rate 
\mbox{$H\sim T^2/\Mpl$} before recombination, but before today, yields
\begin{equation}\label{eq:gNu}
 10^{-8}~\lesssim~\eM^2\,g_X~\lesssim~10^{-7}\;.
\end{equation}
Clearly, this includes the assumption $\mZp\lesssim1\eV$. 
We even focus on the region $\mZp\lesssim10^{-4}\eV$. 
This is crucial to allow kinetic mixing of up to $\chi\lesssim 5\times10^{-8}$
which would be more severely constrained by orders of magnitude for heavier \mZp~\cite{Redondo:2015iea, Gherghetta:2019coi, Fabbrichesi:2020wbt}.
Parametrizing $\mZp=g_X\,\vbar$ we can constrain the size of the effective 
\Up breaking VEV $\vbar$ to
\begin{equation}\label{eq:vbar}
 \vbar=\frac{\mZp}{g_X}~\lesssim~ \eM^2\times 10\times\mathrm{keV}\;.
\end{equation} 
This fixes the necessary hierarchy between the relevant scales of the model to
\begin{equation}\label{eq:xi}
 \xi:=\bar{v}/\vH ~\lesssim~ \eM^2\times4\times10^{-8}\;,
\end{equation}
where $\vH=246\,\GeV$ is the SM Higgs VEV. 
Stabilizing these hierarchies might require tuning in scalar quartic couplings 
which would not change any of our conclusions.
Combining the requirements Eq.~\eqref{eq:couplings} and~\eqref{eq:gNu} and $\chi\lesssim 5\times10^{-8}$ 
we obtain (the lower bound arises from $g_X<1$)
\begin{equation}\label{eq:epsilon}
 10^{-4}~\lesssim~\eM~\lesssim~ 2.5\times10^{-2}\;.
\end{equation}
This implies we automatically, obey constraints arising from violation of
PMNS unitarity \cite{Antusch:2014woa, Fernandez-Martinez:2016lgt}
or direct search bounds~\cite{deGouvea:2015euy,Bryman:2019ssi,Bryman:2019bjg,Bolton:2019pcu}. 
The bounds on $g_X$ and $\eM$ also imply a lower bound
on $\chi\gtrsim10^{-10}$, which would be excluded for $\mZp\gtrsim 10^{-2}\eV$.
This explains why it is not possible to fit the present excess 
for the parameter region considered in \cite{Berbig:2020wve}.
All constraints in the $\eM g_X$-$\chi$ plane are summarized in \Figref{fig:money}.
Note that we assume lower scales for \vP and \vS than in \cite{Berbig:2020wve}
but we do not strive to change the relative hierarchy of \vP and \vS, 
parametrized by the angle 
\begin{equation}
\tan \gamma:=\vP/\vS\;.
\end{equation}
This implies that despite our changes in \mZp, $g_X$ and $\chi$ 
other very characteristic details of this model are
exactly the same as in \cite{Berbig:2020wve}.
This includes key signatures $H\rightarrow\hS\hS$, $H\rightarrow\Zp\Zp$, $H\rightarrow Z\Zp$ and $Z\rightarrow\Zp\hS$ 
whose rates are independent of \mZp, $g_X$, and $\chi$ because they are fixed by Goldstone Boson Equivalence.
These decays contribute to invisible $H$ and $Z$ decays at potentially observable levels, 
which already constrains $s_\gamma\lesssim 0.2$~\cite{Berbig:2020wve}.

Regarding big bang nucleosynthesis (BBN) constraints, 
none of the new light states (\Zp, $h_S$, $N$) was in thermal equilibrium 
with the SM sufficiently before BBN, and between BBN and recombination, as required by BBN~\cite{Fields:2019pfx} and CMB constraints~\cite{Aghanim:2018eyx}.
While thermal abundances of the light states is generated by heavy scalar exchange at temperatures above the electroweak scale,
any such abundance would be depleted by reheating in the SM for example at the QCD phase transition.
Still dangerous is the process $e^+e^-(\nu\bar{\nu})\leftrightarrow N\bar{N}$ 
via $t$-channel $\Phi^\pm$ ($\Phi, A$) exchange.
Absence of this process after QCD (EW) 
epoque requires \mbox{$y\lesssim 6\times 10^{-3(5)} (m_{H^\pm(\Phi)}/100\,\GeV)$}.
Other BBN constraints related to \Zp coupling to neutrinos do not apply here, simply 
because the new gauge interactions become important only after recombination.
The now sizable up-scattering process of Fig.~\ref{fig:nuNee} is cosmologically irrelevant.

The leading direct constraint on the effective neutrino-\Zp coupling arises from allowing unperturbed propagation of SN1987A neutrinos through 
the cosmic neutrino background (C$\nu$B) and implies $\eM^2 g_X\lesssim 5\times10^{-4}$ \cite{Kolb:1987qy} (see also \cite{Konoplich:1988mj}).
Furthermore, our constraint~\Eqref{eq:gNu} already warrants that we are not violating the requirement of free-streaming neutrinos 
during CMB formation \cite{Hannestad:2005ex}.
Laboratory constraints on \mZp and $g_X$ are not very limiting for light mediators (see references collected in~\cite{Berbig:2020wve})
and become even less relevant here as compared to~\cite{Berbig:2020wve} as the 
effective coupling to neutrinos here is smaller by an order of magnitude.

There are strong constraints on dark photon models and kinetic mixing from stellar cooling
if the Higgs mode associated to \Up breaking becomes light \cite{An:2013yua,Redondo:2013lna}.
The relevant Higgs mode in our model is $h_S$ which has 
a sub-keV-scale mass $m_{h_S}\approx\xi\vH\sqrt{2\lambda_S}$,
but it is certainly much heavier than \mZp. 
So the stronger bounds of \cite{An:2013yua} 
(for the Higgsed case), which assume $m_{h_S}\sim\mZp$ do not apply at face value. 
A dedicated analysis in the context of our model would be required, which we
expect to give the leading constraint on $g_X \chi$ directly.
We stress that it is the stellar cooling bounds that matter here,
not the direct detection constraints, 
as the latter can always be avoided if $\Zp$ decays to ($\keV$ energy) 
neutrinos before arriving at the Earth, which is what generically happens 
in our model if $\mZp>2m_\nu$, i.e.\ when \Zp is not a dark matter candidate.

We note there is a parameter region around $\mZp\sim5\times10^{-4}\eV$ and $\chi\sim10^{-9}$
(and for $\mZp<2m_\nu$) where it is not excluded that our \Zp could make up the entirety 
of the Dark Matter (see \cite{Fabbrichesi:2020wbt} and references therein).
For smaller $\mZp$ this possibility is excluded. 
We stress though, that nothing in our resolution of the present XENON1T excess depends on the possibility
of \Zp being the Dark Matter. 
The parameter region $\chi\gtrsim\mathrm{few}\times10^{-9}$ and $\mZp\gtrsim5\times10^{-5}\eV$ will
be probed by ALPS~II \cite{Bahre:2013ywa}.

Taking the expression for the mixing angle \eqref{eq:eM} 
one can show that
\begin{equation}\label{eq:Mbound}
 M~\lesssim~(y/\sqrt{2})\,\eM\,s_\gamma\,\times10\times\keV\;.
\end{equation}
With the above bounds on $y$, $\eM$ and $s_\gamma$
this would imply $M\lesssim0.2\eV$, and $y\vP\lesssim 8\times10^{-3}\eV$. 
Consequently, neutrino masses will not be a small perturbation 
but a substantial ingredient in generating the mixing $\eM$;
including the possibility to lift $M_N$ beyond the above bound.
Hence, generating the mixing -- unlike in \cite{Berbig:2020wve} --
will depend on the details of the neutrino mass generation mechanism.
This might slightly change the valid region of parameters, however, 
it would not change our conclusion that neutrino upscattering to hidden states 
can explain the anomalous excess.
On the other hand, this opens the very exciting possibility that we could investigate 
the neutrino mass generation mechanism
by neutrino electron scattering in XENON1T and complementary experiments, 
just as in the case where the scattering is due to a neutrino magnetic moment.
The exact implications of different neutrino mass generation mechanisms would
have to be studied on a case-by-case basis. We also note that if kinematically allowed by
$M_N$, $N$ might decay to \Zp and neutrinos fast, i.e.\ within the detector volume. 
However, since this is a practically invisible decay this should not leave an observable signature. 

Note that for $m_{\nu,i}>\mZp+m_{\nu,j}$ two- and for higher masses also three-body decays 
of SM neutrinos become possible, depending on the flavor structure of $y$. 
This fact, and the \Zp-mediated 
four-neutrino interaction could substantially modify the cosmic neutrino background.
Sufficiently fast decays would render it mono-generational, 
while the long-range neutrino self-interaction would modify the clustering even leading to
neutrino condensation. In this case the possible coincidence of $\mZp\sim m_\nu\sim T_\mathrm{CMB}$
could become meaningful for the cosmological ``why-now'' problem.

\medskip
In summary, we have outlined a possible explanation for the excess in electronic recoil events
observed in XENON1T. 
In our scenario, the events are caused by inelastic neutrino up-scattering on 
electrons in the detector medium, induced by the standard solar neutrino flux.
Effectively our explanation is based on a light vector mediator that, on the one hand,
has a coupling of SM neutrinos to hidden neutrinos, and on the other hand, 
couples to the electromagnetic current. More specifically, these couplings are understood 
to originate from neutrino mass-mixing and gauge-kinetic mixing between a new 
\Up gauge symmetry with SM hypercharge. 
This model could be discriminated from other explanations of the XENON1T excess 
by the $1/T^2$ recoil-energy dependence of the differential cross section.

We have also presented an explicit gauge invariant, renormalizable and UV-complete model 
which realizes this explanation of the XENON1T excess without 
conflicting with observational constraints.
The model would also lead to long-range neutrino self-interactions that could substantially modify
the appearance of the cosmic neutrino background.
More accessible signatures of the explicit model are new invisible Higgs and $Z$ decays and  
the presence of lepto-philic charged and neutral scalars with masses $\mathcal{O}(100\GeV)$
which all could be searched for at the LHC and future colliders.
Sterile neutrinos are required and they should mix with the SM neutrinos with 
an angle $\eM>10^{-4}$.
The parameter space of the model can further be tested by searches for dark vectors and kinetic mixing,
which is actively pursued, for example at ALPS~II~\cite{Bahre:2013ywa}. 
Leading limits on electron-neutrino scattering are set by TEXONO \cite{Deniz:2010mp}, 
Borexino \cite{Agostini:2018uly} and GEMMA \cite{Beda:2013} experiments \cite{Bilmis:2015lja},
while complementary regions of parameter space will also be probed by electron recoils at 
ongoing reactor neutrino experiments \cite{Lindner:2018kjo,Dent:2019ueq} 
like CONUS \cite{Buck:2020opf}, CONNIE \cite{Aguilar-Arevalo:2016khx} or \textit{$\nu$}-cleus~\cite{Strauss:2017cuu}.

If the excess and our explanation holds up, this may open the exciting possibility of
using neutrino-electron scattering to learn more about the mechanism behind neutrino mass generation
and potentially strong neutrino self-interactions.

\medskip
\begin{acknowledgments}
We thank Xunjie Xu and Manfred Lindner for useful conversations.

\medskip
\textbf{\textit{Note added.---}}%
\textit{After} our paper was submitted to, but before it appeared on the arXiv,
Refs.~\cite{Alonso-Alvarez:2020cdv,Boehm:2020ltd,Fornal:2020npv} appeared which also treat the XENON1T excess
(See also Ref.~\cite{Bell:2020bes} which was submitted before our paper appeared).
Especially \cite{Boehm:2020ltd} also considers neutrino-electron scattering by a light vector mediator, despite 
in a somewhat effective picture without complete model.
Our results are consistent where overlapping, but we stress that mediator masses 
heavier than $\mZp\gtrsim 1\mathrm{eV}$ are excluded by Dark Photon constraints; 
which unavoidably brings long-range neutrino self-interaction into focus.
\end{acknowledgments}

\bibliographystyle{utphys}
\bibliography{Orbifold}

\end{document}